# Accuracy of extracted optical conductivity by Kramer-Kronig analysis from reflectivity spectrum

JEONG WOO HAN[1]

[1]*Department of Physics Education, Chonnam National University, Gwangju 61186, South Korea*
*jwhan@chonnam.edu*

**Abstract:** Kramers-Kronig (KK) analysis has been widely used to extract the optical conductivity spectrum from a broad range of reflectance spectrum obtained from far-infrared to ultraviolet frequency ranges. In this study, we present how measurement uncertainty in the reflectivity spectrum affects the extracted optical conductivity spectrum obtained through KK analysis. We consider realistic uncertainties that can easily occur in reflectance measurement environments: (1) a rigid shift of the absolute reflectance in the whole measurement frequency window, and (2) a linear decrement of reflectance with increasing frequency. Our investigation reveals that the reliability of the extracted optical conductivity spectrum, especially in the lower-frequency range, should be carefully addressed, particularly when the reflectance is above approximately 95 %.

## 1. INTRODUCTION

Optical spectroscopy enables non-destructive investigation of various materials' optical properties without physical contact for which various types of optical spectroscopic techniques have been developed [1-7]. Fourier transform infrared spectroscopy (FT-IR) is an efficient measurement technique for the investigation of linear optical properties [8-13]. For FT-IR, we can obtain transmittance and reflectance spectra across a broad frequency range—from far-infrared to mid-infrared frequency range. Unlike other optical measurement techniques, FT-IR measures the intensity of transmitted and reflected electromagnetic waves. This fact suggests that the phase information of incident electromagnetic waves, mainly determined by the optical traveling distance, has little influence on the experimentally obtained transmittance and/or reflectance from FT-IR. Namely, sample position plays a minor role in the experiment data of FT-IR. This advantage is particularly pronounced in reflectance measurements in FT-IR compared to other spectroscopic techniques. For example, terahertz time-domain spectroscopy directly measures transmitted and/or reflected electric fields, suggesting that the position between the sample and the reference should be ensured less than a sub-micrometer scale to obtain the accurate measurement of response functions, e.g., optical conductivity, dielectric constant, and refractive index [14-16]. However, FT-IR alone cannot directly provide optical response functions, which is one of the biggest drawbacks of FT-IR.

The Kramers-Kronig (KK) relations are the mathematical formulation enabling the connection between the real and the imaginary parts of any complex functions [17-19]. This connection directly derives from the principle of causality, which allows us to obtain the optical response function from the broad range of transmittance and/or reflectance spectrum. Therefore, by applying KK analysis to experimental data obtained from FT-IR, one can obtain the response functions, which makes KK relations a powerful analysis methodology. However, there was no qualitative analysis of how the error in the experimentally measured spectrum affects the obtained response function.

In our study, we investigate the influences of the uncertainty on the optical response function extracted through Kramers-Kronig (KK) analysis when the experimentally obtained reflectance data is inaccurate. Specifically, we focus on the optical conductivity as the optical response function. To reproduce the realistic uncertainty of the reflection-type measurement, we arbitrarily down-shift the reflectance spectrum for the whole measurement frequency range with a constant value. This type of measurement error frequently arises in reflectance-type measurement due to mismatches of the reflectance areas between the sample and the reference. Additionally, we consider an unexpected decline in the reflectance spectrum with increasing frequency. Such a decline can occur due to the non-flat morphology of the sample surface with certain roughness thereby inducing diffuse reflection. This fact leads to the higher (lower) detection efficiency on the lower (higher) frequency region.

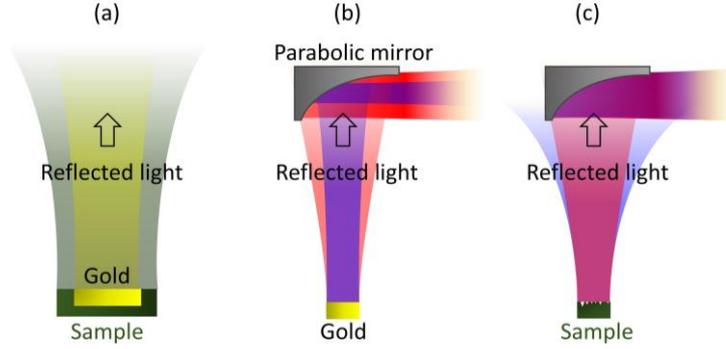

**Fig. 1.** (a) Sketch of different areas of reflected light between the sample and the reference (gold). Frequency-dependent divergence of the reflected light from the gold (b) and the sample (c). Blue and red lights denote the higher and the lower frequencies of light. Diffuse reflection occurs at the sample with non-flat morphology with certain roughness, leading to the fact that the higher frequency of light is more divergence.

## 2. RESULTS AND DISCUSSION

Reflectance in normal incident geometry can be expressed by using frequency-dependent Fresnel's reflectance coefficient $r(\omega)$, as:

$$r(\omega) = \frac{n(\omega) - 1 + ik(\omega)}{n(\omega) + 1 + ik(\omega)} \quad (1)$$

where $n(\omega)$ and $k(\omega)$ are the frequency-dependent real and imaginary parts of the complex refractive index of the sample. The experimental data obtained from FT-IR is the intensity of the reflected incident light. The intensity ratio between the sample and the reference is corresponding to the reflectance $R(\omega)$, which is the square of $r(\omega)$, as:

$$R(\omega) = |r(\omega)|^2 \quad (2)$$

As can be seen in Eq. (2), phase information $\theta(\omega)$ of $r(\omega)$ is canceled in the experimental data such that it is impossible to directly retrieve the two independent variables of $n(\omega)$ and $k(\omega)$. However, this missing information can be restored through KK analysis, which is given as [20]:

$$\theta(\omega) = -\frac{2\omega}{\pi} P \int_0^\infty \frac{\ln|r(\omega')|}{\omega'^2 - \omega^2} d\omega' \quad (3)$$

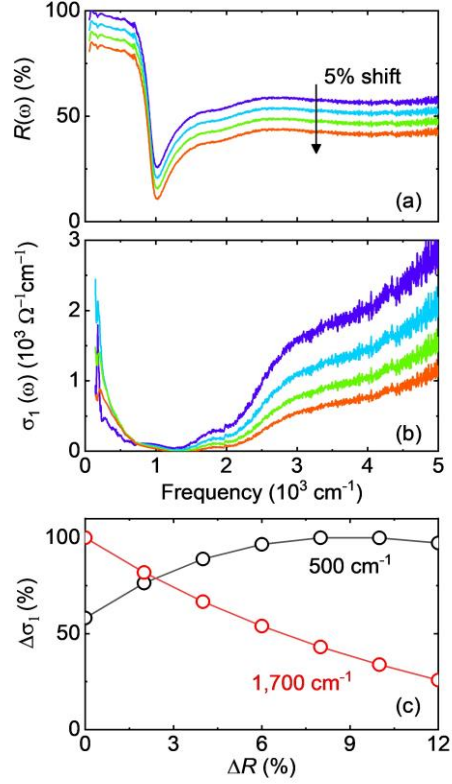

**Fig. 2.** (a) The reflectance spectrum of SrMnSb$_2$, took from Ref. [25]. Reflectance is down-shifted with the interval of 5%. (b) Converted real part of optical conductivity spectrum $\sigma_1(\omega)$ through KK analysis. (c) Change of $\sigma_1$ extracted from 500 cm$^{-1}$ and 1,700 cm$^{-1}$.

Here, $P$ denotes the Cauchy principal value. However, in real-world measurements, the integration range cannot extend from 0 to infinity. To avoid this limitation, a common approach is the extrapolate technique applied to the high-frequency range. Specifically, the extrapolation method that ensures continuity with the last measurement value of $R(\omega)$ has been commonly adopted, as [21]:

$$R(\omega) = R_L \left(\frac{\omega}{\omega_L}\right)^{-\alpha} \quad (4)$$

where $R_L(\omega)$ and $\omega_L$ represent the last value of $R(\omega)$ and the highest frequency value in the measurement data, respectively. The parameter $\alpha$ is a free variable. We confirmed that $\alpha$ does not give a significant change on the optical conductivity spectrum, provided that $\omega_L$ is around 5 eV, which can be accessible using commercialized ellipsometry [22,23]. By combining $R(\omega)$ and $\theta(\omega)$, one can retrieve $n(\omega)$ and $k(\omega)$ by using below relations, as:

$$n(\omega) = \frac{1 - R(\omega)}{1 + R(\omega) - 2\sqrt{R(\omega)}\cos\theta(\omega)} \quad (5)$$

$$k(\omega) = \frac{2\sqrt{R(\omega)}\sin\theta(\omega)}{1 + R(\omega) - 2\sqrt{R(\omega)}\cos\theta(\omega)} \quad (6)$$

The complex optical conductivity spectrum σ(ω)= $\sigma_1(\omega)+i\sigma_2(\omega)$ can be converted from $n(\omega)$ and $k(\omega)$ by using below relation, as [24]:

$$\sigma_1(\omega) = \frac{n(\omega)k(\omega)\omega}{2\pi} \qquad (7)$$

$$\sigma_2(\omega) = (1 - n^2(\omega) - k^2(\omega))\frac{\omega}{4\pi} \qquad (8)$$

Note that we focus on $\sigma_1(\omega)$ as the response function in this entire study.

To accurately measure the absolute value of reflectance, the reflective areas between the sample and the reference should be identical. To achieve this, the gold deposition technique has been employed, where gold is deposited on the sample surface to use the sample itself as a reference [25-27]. However, this method has the significant drawback that the sample can no longer be used. Therefore, obtaining accurate reflectance values while preserving the sample remains a very challenging task. Figure 1(a) depicts the mismatch of reflective areas. As can be seen, the absolute quantities of reflectance can be overestimated and/or underestimated by the different reflective areas between the sample and the reference. Note that the perfect reflectors such as gold have been usually employed as the reference.

To reproduce this measurement inaccuracy, we down-shifted rigidly the entire reflectance values up to 15 % with the interval of 5 % as shown in Fig. 2(a). Note that we took reflectance data of SrMnSb$_2$ from Ref. [28] and the original data was denoted by blue. Figure 2(b) shows the extracted $\sigma_1(\omega)$. The dip structure $\omega_p$ observed around 1,000 cm$^{-1}$ in the reflectance spectrum corresponds to the effective plasma frequency. Below $\omega_p$, the free-carrier response serves as the dominant contribution to the reflectance spectrum, and the rapid increase of $\sigma_1(\omega)$ observed in low-frequency region is a typical signature of free-carrier dynamics in the reflectance spectrum.

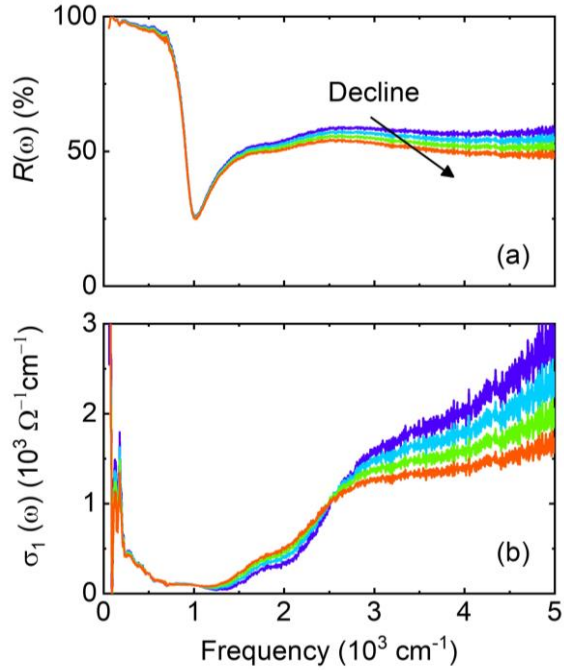

**Fig. 3.** (a) The reflectance spectrum of SrMnSb$_2$, took from Ref. [25]. Reflectance is artificially adjusted to have lower reflectance at higher frequency. At 5,000 cm$^{-1}$, reflectance data was set to decrease by approximately 3% for each data set, denoted by different color codes. (b) Converted real part of optical conductivity spectrum $\sigma_1(\omega)$ through KK analysis.

One can monitor that the free-carrier response on $\sigma_1(\omega)$ generally increases whereas overall spectral weight of $\sigma_1(\omega)$ decreases as the reflectance is reduced. This feature is clearly confirmed in Fig. 2(c), which shows the change of $\sigma_1$ at 500 cm$^{-1}$ and 1,700 cm$^{-1}$ as a function of the percentage of the down-shifted reflectance $\Delta R$. When the reflectance is reduced about 3% from the original reflectance, $\sigma_1$ at 500 cm$^{-1}$ increases about 15 %. However, when $\Delta R$ is above 6 %, less than 5% change in $\sigma_1$ at 500 cm$^{-1}$ is observed. By examining the multiple reflectance spectra reported in various samples, we can conclude that the free-carrier response in obtained $\sigma_1$ is very sensitive even to small measurement errors in the reflectance especially when the free-carriers nearly reflect the incident light thus leading to very high reflectance (~ 95%). In other words, when analyzing free-carrier dynamics with high reflectance based on the KK analysis, it is very crucial to measure the accurate absolute value of reflectance. For $\Delta\sigma_1$ at 1,700 cm$^{-1}$, the monotonically decreasing behavior is observed upon increasing $\Delta R$.

In actual reflection measurements, the sample surface usually has a non-flat morphology with certain roughness. Therefore, the reflected beam between the sample and the reference is detected with different detection efficiency, resulting from diffuse reflection, which becomes stronger as the frequency of the incident light increases. Additionally, since the size of collection optics such as parabolic mirrors is realistically limited, the amount of detected light decreases as the frequency increases. Consequently, this fact leads to a decline profile in the reflectance spectrum with increasing frequency. Figure 1(b) and (c) describe the reflections between the reference (gold) and the sample with non-flat morphology of surface with certain roughness, respectively. As the reference has a flat surface and clean roughness, the reflected light with higher frequency (blue color) is less divergent compared to the corresponding light with lower frequency (red color), as shown in Fig. 1(b). However, for the sample with certain roughness, diffuse reflection is expected, leading to the opposite situation where the higher frequency is less collected (see Fig. 1(c)).

To reproduce this decline reflectance profile, we artificially adjust to have lower reflectance as increasing frequency with the linearly decrement trend. Note that reflectance was set to have lower value with the interval of about 3 % for each data set at frequency of 5000 cm$^{-1}$ (see Fig. 3(a)). As the same to Fig. 2, we took the original reflectance data from Ref. [28]. Figure 3(b) shows the corresponding converted $\sigma_1(\omega)$ through KK analysis. As can be seen, the errors in $\sigma_1(\omega)$ become more pronounced with increasing frequency. This trend can be naturally understood by the fact that deviations from the original reflectance spectrum more accumulate by increasing frequency. In other words, when the sample morphology is not a flat and even the surface roughness is not fine compared to the wavelength of the incident beam, the reliability of $\sigma_1(\omega)$ of the high frequency region should be carefully addressed.

## 3. CONCLUSION

In conclusion, we have conducted quantitative investigations into the influence of measurement uncertainty on reflectance measurements and their impact on the converted optical conductivity through Kramer-Kronig analysis. We considered realistic uncertainty errors that can easily occur in actual reflectance measurement environments: (1) inaccuracies in overall reflectance values stemming from different reflection areas between the sample and the reference, and (2) the decline in reflectance due to the non-flat morphology of the sample surface with certain roughness. Our investigation revealed that (1) can lead to overall inaccuracies in optical conductivity, especially when the reflectance is very high (i.e., above 95%). The error type of (2) can cause inaccuracies particularly in the high-frequency region, while the optical conductivity in the lower frequency range is minimally affected. We believe that this study will provide valuable guidelines for scientists and engineers working with optical spectroscopic tools and Kramer-Kronig analysis.